\documentstyle[epsfig,12pt]{article}

\begin{document}

\mbox{ } \\[-1cm]
\mbox{ }\hfill KEK--TH--698\\%[1cm] 
\mbox{ }\hfill KIAS--P00028\\%[-1mm] 
\mbox{ }\hfill OCHA--PP--160\\%[-1mm] 
\mbox{ }\hfill hep--ph/0005313\\%[-1mm] 
\mbox{ }\hfill \today\\%[-1mm]

\begin{center}
{\Large\bf Measuring the Higgs CP Property through Top Quark \\[2mm]
            Pair Production at Photon Linear Colliders}\\[1.4cm]
Eri Asakawa$^1$, S.Y. Choi$^2$, Kaoru Hagiwara$^3$ and Jae Sik Lee$^4$
\end{center}

% \hskip 0.2cm  
%
\begin{center}
$^1${\it Graduate School of Humanities and Sciences, Ochanomizu University,\\ 
     1--1 Otsuka 2--chome, Bunkyo, Tokyo 112-8610, Japan}\\[2mm]
$^2${\it Department of Physics, Chonbuk National University, 
     Chonju 561--756, Korea} \\[2mm] 
$^3${\it Theory Group, KEK, Tsukuba, Ibaraki 305-0801, Japan}\\[2mm]
$^4${\it Korea Institute for Advanced Study, Seoul 130--012, Korea}
\end{center}

\bigskip
\bigskip 
\hskip 2cm

\begin{abstract}
We present a model--independent study of the effects of a neutral Higgs 
boson without definite CP--parity in the process $\gamma\gamma
\rightarrow t\bar{t}$ around the mass pole of the Higgs boson.
Near the resonance pole, the interference between the Higgs--exchange and 
the continuum amplitudes can be sizable if the photon beams are polarized 
and helicities of the top and anti--top quarks are measured. 
Study of these interference effects enables one to determine 
the CP property of the Higgs boson completely. An example of the complete 
determination is demonstrated in the context of the minimal supersymmetric 
standard model.
\end{abstract}
%

%\pacs{PACS number(s): 11.30.Er, 12.60.Jv, 13.10.+q}

\renewcommand{\thefootnote}{\alph{footnote}}

\newpage

%%%%%%%%%%%%%%%%%%%%%%%%%%%%%%
\section{Introduction}
\label{sec:introduction}
%%%%%%%%%%%%%%%%%%%%%%%%%%%%%%

Search for Higgs bosons and precise measurements of their 
properties such as their masses, the decay widths and the decay 
branching ratios \cite{Hunter} are among the most important subjects in our
study of the electroweak symmetry breaking.  
In the standard model (SM), only one physical neutral Higgs boson appears
and its couplings to all the massive particles are uniquely determined.
On the other hand, models with multiple Higgs doublets have 
neutral Higgs bosons of definite CP parity as well as charged Higgs bosons, 
if CP is a good symmetry. If CP is not a good symmetry
of the symmetry breaking physics, these neutral Higgs bosons
do not necessarily carry definite CP parity.\\ 

The CP--violating interactions beyond the Kobayashi--Maskawa mechanism 
and their consequences at high--energy colliders have been intensively
studied. This is motivated in part by the search of an efficient 
mechanisms of generating the cosmological baryon asymmetry at the 
electroweak scale \cite{EWBG}. An extended Higgs--boson sector, as predicted 
by many extensions of the SM such as the minimal 
supersymmetric SM (MSSM), can provide such CP--violating interactions 
in a natural way. One attractive scenario is to make use of explicit 
CP violations in the MSSM Higgs sector \cite{EXCP,CDL}
which are induced through loop corrections
with complex supersymmetric parameters in the mass matrices of the
third generation squarks. Such interactions cause mixings among 
CP--even and CP--odd Higgs bosons.\\

It is of great interest to examine the possibility of studying these 
Higgs bosons in detail in CP non--invariant theories. A two--photon 
collision option \cite{ggoriginal} of the future linear $e^+ e^-$ 
colliders offers one of the ideal places to look for such Higgs signals 
\cite{ggcol}. There, neutral Higgs bosons can be produced via loop 
diagrams of charged particles.  If the Higgs boson is lighter than about 
140 GeV,  its two-photon decay width can be 
measured accurately by looking for its main decay mode, which is 
usually the $b\overline{b}$ mode \cite{OTW}.  
When the Higgs boson is heavy, the processes $\gamma\gamma 
\rightarrow W^+W^-, ZZ$ and $t\bar{t}$ can be useful to detect 
its signal \cite{Jikia}. In theories with weakly coupled Higgs sector,
such as the MSSM, heavier Higgs bosons have suppressed branching fractions
to the $WW/ZZ$ modes \cite{HDecay}. Furthermore, CP-odd Higgs bosons do not 
have the $ZZ$ decay modes in the tree level. 
On the other hand, the $t\overline{t}$ decay mode 
can be significant irrespective of the CP property
of the Higgs boson $\phi$. In this case, it is expected that
the heavy Higgs boson contributes to the process $\gamma\gamma\rightarrow 
t\overline{t}$ significantly around its mass pole.\\  

The $s$--channel resonant amplitude of $\gamma\gamma\rightarrow
\phi^{(*)}\rightarrow t\overline{t}$ can interfere with the tree--level
$t$-- and $u$--channel continuum amplitudes, if the resonant and the 
continuum amplitudes have comparable magnitudes near the resonance pole
\cite{AKSW}.  
This often happens for heavy Higgs bosons where both the peak resonant
amplitudes and their total decay widths are large enough to make
the interference effects significant.
In this article, we study the contribution of a heavy Higgs boson 
without definite CP--parity to the process $\gamma\gamma \rightarrow 
t\overline{t}$ and describe an efficient method to determine its
CP property completely by making use of the photon and 
$t\bar{t}$ polarizations;
the procedure is crucially based upon the interference effects among 
various helicity amplitudes. We also demonstrate the feasibility of 
determining the heavy Higgs--boson contribution in 
the context of the MSSM.\\

The remainder of this article is organized as follows. 
In Sect.~\ref{sec:helicity amplitude}, 
the helicity amplitudes of the process $\gamma\gamma\rightarrow
t\bar{t}$ are calculated with model--independent parametrizations
of the couplings of the Higgs boson $\phi$ to a photon pair and a
top--quark pair. In Sect.~\ref{sec:polarized cross section} 
we present all the observables constructed 
by use of photon polarizations and by measuring the helicities of the
final top and anti--top quarks. Section~\ref{sec:complete diagnosis}
gives the description of the procedure to determine 
the CP property of the Higgs boson completely.
In Sect.~\ref{sec:polarized photon beam}
we study the properties of polarized photon beams through the
Compton laser backscattering. In Sect.~\ref{sec:example} we demonstrate 
the complete determination of the CP property of the heaviest 
neutral Higgs boson in the MSSM.  Finally, we give conclusions in 
Sect.~\ref{sec:conclusion}.

%%%%%%%%%%%%%%%%%%%%%%%%%%%%%%%%%%%%%%%%%%%%%%%%%%%%%%%%%%%%%%%%%
\section{Helicity amplitudes}
\label{sec:helicity amplitude}
%%%%%%%%%%%%%%%%%%%%%%%%%%%%%%%%%%%%%%%%%%%%%%%%%%%%%%%%%%%%%%%%%

In CP non--invariant theories, the lowest--dimensional interaction 
of the Higgs boson $\phi$ with a top--quark pair can be described 
in a model--independent way
by the vertex:  
\begin{eqnarray}
{\cal V}_{\phi t\overline{t}}=-ie\, \frac{m_t}{m_W}\,
           \left(S_t+i\gamma_5 P_t\right) \,,
\label{vtx_t}
\end{eqnarray}
and the loop--induced interaction of the Higgs boson with a photon pair
is parameterized in a model--independent form as follows:
\begin{eqnarray}
{\cal V}_{\gamma\gamma\phi}&=&\sqrt{s}\frac{\alpha}{4\pi}\,
   \left\{S_\gamma(s) \left[\epsilon_1\cdot\epsilon_2
                    -\frac{2}{s}(\epsilon_1\cdot k_2)(\epsilon_2\cdot k_1)
		 \right]
         -P_\gamma(s) \frac{2}{s} \left\langle \epsilon_1\epsilon_2 k_1 k_2
                 \right \rangle \right\}\,,
\label{vtx_r}
\end{eqnarray}
where the form factors depend on the c.m. energy squared $s$ of two colliding 
photons,  $\epsilon_{1,2}$ stand for the wave vectors of the two photons,
$k_{1,2}$ are the four--momenta of the two photons, and
\begin{eqnarray}
\langle\epsilon_1\epsilon_2 k_1 k_2\rangle = 
       \epsilon_{\mu\nu\alpha\beta}\, \epsilon_1^\mu\epsilon_2^\nu 
       k_1^\alpha k_2^\beta\,, 
\end{eqnarray}
with $\epsilon_{0123}=1$.
Since we are interested in 
the $s$--channel virtual (and real) Higgs--boson exchange, for the sake of 
consistency the $s$--dependence of the form factors $S_\gamma$ and 
$P_\gamma$ is exhibited explicitly in Eq.~(\ref{vtx_r}). We note that
a simultaneous presence of $\{S_\gamma, P_\gamma\}$ or/and
$\{S_t,P_t\}$ implies CP non--invariance of the theory.\\

In the two--photon c.m coordinate system with $\vec{k}_1$ along 
the positive $z$ direction and $\vec{k}_2$ along the 
negative $z$ direction, the wave vectors $\epsilon_{1,2}$ of two photons 
are given by
\begin{eqnarray}
\epsilon_1^\mu (\lambda)=\epsilon^{\mu*}_2(\lambda)
                =\frac{-\lambda}{\sqrt{2}}\,\bigg(0,1,i\lambda,0\bigg)\,.
\label{epsil}
\end{eqnarray}
where $\lambda=\pm 1$ denote the right and left photon helicities,
respectively. Using Eqs.~(\ref{vtx_t}), (\ref{vtx_r}) and ~(\ref{epsil}) 
one can derive the explicit forms of the helicity amplitudes for 
the process $\gamma\gamma\rightarrow t\overline{t}$ which consist of two 
parts; the s--channel Higgs--boson exchange and the tree--level 
continuum contributions:\\

\noindent
{\bf (i) Higgs--exchange contribution}:\\
\begin{eqnarray}
{\cal M}_\phi^{(\lambda_1\lambda_2:\sigma\overline{\sigma})}
   = \frac{e\,\alpha\, m_t}{4\pi m_W}
     \frac{s}{s-m_\phi^2+im_\phi\Gamma_\phi} 
     \left[\,S_\gamma(s) +i\lambda_1 P_\gamma(s)\right]
     \left[\,\sigma\beta S_t-i P_t\right]
     \delta_{\lambda_1,\lambda_2}\delta_{\sigma\overline\sigma}\,.
\label{mephi}
\end{eqnarray}\\

\noindent
{\bf (ii) Tree--level continuum contribution}:\\
\begin{eqnarray}
&& {\cal M}_{\rm cont}^{(\lambda_1\lambda_2:\sigma\overline{\sigma})}
    =
\frac{4\pi\alpha Q_t^2}{1-\beta^2\cos^2\Theta}
\left\{\frac{4m_t}{\sqrt{s}}(\lambda_1+\sigma\beta)\,
\delta_{\lambda_1\lambda_2}\delta_{\sigma\overline\sigma}
\right. \nonumber \\
&& 
\left.
~~~~~~~~~~~~~~
-\frac{4m_t}{\sqrt{s}}\sigma\beta\sin^2\Theta\,
\delta_{\lambda_1,-\lambda_2}\delta_{\sigma\overline\sigma}
-2\beta(\cos\Theta+\lambda_1\sigma)\sin\Theta\,
\delta_{\lambda_1,-\lambda_2}\delta_{\sigma,-\overline\sigma}
\right\}\,,
\label{mecont}
\end{eqnarray}
where $\Theta$ is the scattering angle of the top quark with respect to
the positive $z$ direction, $\beta=\sqrt{1-4m_t^2/s}$, 
$\lambda_1 =\pm 1$ ($\lambda_2=\pm 1$) denote the right and left helicities
of the incident photon with its three--momentum along the positive (negative)
direction, respectively, in units of $\hbar$, and
$\sigma=\pm 1$ ($\overline\sigma=\pm 1$) denote the right and left 
helicities of $t$ ($\overline{t}$), respectively, in units of 
$\hbar/2$. There 
exists the interference between the Higgs--boson and the continuum 
contributions for equal $t$ and $\bar{t}$ helicities 
which can be used to observe the
CP property of the Higgs boson. On the other hand, the continuum contribution
with opposite $t$ and $\bar{t}$ helicities dominates the $\gamma\gamma
\rightarrow t\bar{t}$ events at high energies so that it is important to
distinguish events with equal $t$ and $\bar{t}$ helicities from those
with opposite $t$ and $\bar{t}$ helicities. With the above points in mind, 
we concentrate on the helicity amplitudes of equal $t$ and $\bar{t}$ 
helicities in the present work.\\
 
When the helicities of the top and anti--top quarks are equal, i.e.
$\sigma=\overline\sigma$, the helicity amplitudes for the process
$\gamma\gamma\rightarrow t\overline{t}$ can be rewritten in the following
simple form
\begin{eqnarray}
{\cal M}^{(\lambda\lambda:\sigma\sigma)}
   &=&
A_{\rm cont}(s)(\lambda+\sigma\beta)+
A_\phi(s) \left[S_\gamma+i\lambda P_\gamma\right]
\left[\sigma\beta S_t-i P_t\right]\,,
\nonumber \\
{\cal M}^{(\lambda,-\lambda:\sigma\sigma)}
   &=&
-A_{\rm cont}(s)\sigma\beta\sin^2\Theta\,,
\label{me}
\end{eqnarray}
where for the sake of brevity the explicit $s$--dependence of the form 
factors $S_\gamma$ and $P_\gamma$ is not denoted and two $s$--dependent
functions $A_{\rm cont}(s)$ and $A_\phi(s)$ are introduced:
\begin{eqnarray}
A_{\rm cont}(s)=
\frac{16\pi\alpha\,Q_t^2\,m_t}{\sqrt{s}\,(1-\beta^2\cos^2\Theta)}
\,,\qquad
A_\phi(s)= \frac{e\,\alpha}{4\pi}\frac{m_t}{m_W}\,D_\phi(s)\,.
\end{eqnarray}
Here $D_{\phi}(s)$ is the $s$--channel propagator of the Higgs boson
\begin{eqnarray}
D_\phi(s)=\frac{s}{s-m_\phi^2+im_\phi\Gamma_\phi}\,.
\end{eqnarray}
A few important comments concerning the helicity amplitudes (\ref{me}) 
are in order:
\begin{itemize}
\item Only the amplitudes with equal photon helicities
      have contributions from the Higgs--boson exchange.
\item The spin--0 Higgs--boson contributions are independent of the scattering
      angle $\Theta$.
\item The $s$--dependent form factors $S_\gamma$ and $P_\gamma$ as well as
      $A_\phi$ are in general complex while the other form factors
      are real in the leading order.
\item The continuum part is CP--preserving while the Higgs--exchange 
      part can be CP--violating if the 
      `scalar' (CP--even) form factors, $S_\gamma$ and $S_t$, and
      `pseudoscalar' (CP--odd) form factors, $P_\gamma$ and $P_t$,
      are present simultaneously.
\end{itemize}
%

%%%%%%%%%%%%%%%%%%%%%%%%%%%%%%%%%%%%%%%%%%%%%%%%
\section{Polarized cross sections}
\label{sec:polarized cross section}
%%%%%%%%%%%%%%%%%%%%%%%%%%%%%%%%%%%%%%%%%%%%%%%%

%%%%%%%%%%%%%%%%%%%%%%%%%%%%%%%%%%%%%%%%%%%%%%%%%%%%%%%%%%%%%%%%%%%%%%%%
\subsection{Equal photon helicities and top--quark helicities}
%%%%%%%%%%%%%%%%%%%%%%%%%%%%%%%%%%%%%%%%%%%%%%%%%%%%%%%%%%%%%%%%%%%%%%%%

In this section, we investigate what physics information on the 
Higgs--boson contribution can be extracted 
from the cross section with equal photon and top--quark helicities. 
First of all, we can construct four independent squared amplitudes:
\begin{eqnarray}
&&\left|{\cal M}^{(++:++)}\right|^2=\overline{\left|{\cal M}\right|^2_0}
\,\bigg[1+{\cal A}_0 +{\cal A}_1-(1+\beta)({\cal A}_2-{\cal A}_3)\bigg]
\,,\nonumber \\
&&\left|{\cal M}^{(--:--)}\right|^2=\overline{\left|{\cal M}\right|^2_0}
\,\bigg[1+{\cal A}_0 -{\cal A}_1+(1+\beta)({\cal A}_2-{\cal A}_3)\bigg]
\,,\nonumber \\
&&\left|{\cal M}^{(++:--)}\right|^2=\overline{\left|{\cal M}\right|^2_0}
\,\bigg[1-{\cal A}_0 +{\cal A}_1+(1-\beta)({\cal A}_2+{\cal A}_3)\bigg]
\,,\nonumber \\
&&\left|{\cal M}^{(--:++)}\right|^2=\overline{\left|{\cal M}\right|^2_0}
\,\bigg[1-{\cal A}_0 -{\cal A}_1-(1-\beta)({\cal A}_2+{\cal A}_3)\bigg]
\,,
\end{eqnarray}
where $\overline{\left|{\cal M}\right|^2_0}$ is the unpolarized
squared amplitude, i.e. the average
\begin{eqnarray}
\overline{\left|{\cal M}\right|^2_0} = \frac{1}{4}
         \left[\left|{\cal M}^{(++:++)}\right|^2
               +\left|{\cal M}^{(--:--)}\right|^2
               +\left|{\cal M}^{(++:--)}\right|^2
               +\left|{\cal M}^{(--:++)}\right|^2\right]\,.
\end{eqnarray}
The explicit form (\ref{me}) of the helicity amplitudes then leads to 
the following expressions for $\overline{\left|{\cal M}\right|^2_0}$ 
and the quantities ${\cal A}_i$ ($i=0,1,2,3$):
\begin{eqnarray}
\overline{\left|{\cal M}\right|^2_0} 
  &=&(1+\beta^2) A_{\rm cont}^2
    +(\beta^2 S_t^2+P_t^2)(\left|S_\gamma\right|^2 +\left|P_\gamma\right|^2)
     \left|A_\phi\right|^2 \nonumber \\
  &&+2A_{\rm cont}\left[ \vphantom{ {\cal M}^{(++:++)} }
     \beta^2 S_t\,{\cal R}(A_\phi S_\gamma)
    +P_t{\cal R}(A_\phi P_\gamma)\right]\,,\nonumber\\
{\cal A}_0
  &=&2\beta A_{\rm cont}\bigg\{A_{\rm cont}+
     \left[S_t\,{\cal R}(A_\phi S_\gamma)
          +P_t\,{\cal R}(A_\phi P_\gamma)\right]\bigg\}
     \,\bigg/\overline{\left|{\cal M}\right|^2_0}\,,\nonumber \\
{\cal A}_1
  &=&2\left|A_\phi\right|^2\,\bigg\{(\beta^2 S_t^2+P_t^2)
      \,{\cal I}(S_\gamma P^*_\gamma)\bigg\}\,
      \bigg/\overline{\left|{\cal M}\right|^2_0}\,,\nonumber \\
{\cal A}_2
  &=&2\beta A_{\rm cont}\,\bigg\{ S_t\,{\cal I}(A_\phi P_\gamma)\bigg\}
      \,\bigg/\overline{\left|{\cal M}\right|^2_0}\,,\nonumber \\
{\cal A}_3
  &=&2 A_{\rm cont}\, \bigg\{P_t\,{\cal I}(A_\phi S_\gamma)\bigg\}
     \,\bigg/\overline{\left|{\cal M}\right|^2_0}\,,
\end{eqnarray}
where $\overline{\left|{\cal M}\right|^2_0}$ and ${\cal A}_0$ are 
CP--even, but the three asymmetries ${\cal A}_{1,2,3}$ are 
CP-odd, that is to say, they can be non--vanishing only in CP non--invariant
theories. More explicitly, we can exploit the CP--odd combinations
\begin{eqnarray}
% \left\{\left[\left|{\cal M}^{(++:++)}\right|^2
%            -\left|{\cal M}^{(++:--)}\right|^2\right]
%-\left[\left|{\cal M}^{(--:--)}\right|^2
%      -\left|{\cal M}^{(--:++)}\right|\right]^2\right\}
%     \,\bigg/{4\overline{\left|{\cal M}\right|^2_0}}
\sum_{\lambda}\lambda 
 \left\{\left|{\cal M}^{(++:\lambda\lambda)}\right|^2
       +\left|{\cal M}^{(--:\lambda\lambda)}\right|^2\right\}
      \,\bigg/{4\overline{\left|{\cal M}\right|^2_0}}
  &=& -{\cal A}_2 +\beta{\cal A}_3\,, \nonumber \\
% \left\{\left[\left|{\cal M}^{(++:++)}\right|^2
%       -\left|{\cal M}^{(--:++)}\right|^2\right]
%-\left[\left|{\cal M}^{(--:--)}\right|^2
%      -\left|{\cal M}^{(++:--)}\right|\right]^2\right\}
%      \,\bigg/{4\overline{\left|{\cal M}\right|^2_0}}
\sum_{\sigma}\sigma 
 \left\{\left|{\cal M}^{(\sigma\sigma:++)}\right|^2
       +\left|{\cal M}^{(\sigma\sigma:--)}\right|^2\right\}
      \,\bigg/{4\overline{\left|{\cal M}\right|^2_0}}
  &=& {\cal A}_1 -\beta{\cal A}_2 +{\cal A}_3\,,
\end{eqnarray}
in extracting the CP--odd asymmetries ${\cal A}_{1,2,3}$.
It is however clear that we need to exploit more observables to 
determine all the form factors, $S_\gamma$, $S_t$, $P_\gamma$ and 
$P_t$ completely. This can be
done by using linear photon polarization as well as circular
photon polarization as shown in the next section. \\

%%%%%%%%%%%%%%%%%%%%%%%%%%%%%%%%%%%%%%%%%%%%%%%%%%%%%%%%%%%%%%%%
\subsection{Two photon spin correlations}
%%%%%%%%%%%%%%%%%%%%%%%%%%%%%%%%%%%%%%%%%%%%%%%%%%%%%%%%%%%%%%%%

Taking into account the general polarization configuration of two
photon beams and taking the sum over the final polarization
configuration with equal $t$ and $\bar{t}$ helicities, we obtain 
the polarized squared amplitude as 
\begin{eqnarray}
\overline{\left|{\cal M}\right|^2}&=&
\overline{\left|{\cal M}\right|^2_0}^{\,\prime}\,\,
\bigg\{    \left(1+\zeta_2\tilde{\zeta}_2\right)
+{\cal B}_1\left(\zeta_2+\tilde{\zeta}_2\right)
+{\cal B}_2\left(\zeta_1\tilde{\zeta}_3+\zeta_3\tilde{\zeta}_1\right)
-{\cal B}_3\left(\zeta_1\tilde{\zeta}_1-\zeta_3\tilde{\zeta}_3\right)
\nonumber \\ &&
~~~~~~~~~
+\sin^2\Theta\left[-{\cal C}_0\left(\zeta_2\tilde{\zeta}_2
-\zeta_3\tilde{\zeta}_3 \right)
+{\cal C}_1\left(\zeta_1+\tilde{\zeta}_1\right)
+{\cal C}_2\left(\zeta_3+\tilde{\zeta}_3\right)
\right. \nonumber \\ && \left.
~~~~~~~~~~~~~~~~~~~~~
+{\cal C}_3\left(\zeta_1\tilde{\zeta}_2+\zeta_2\tilde{\zeta}_1\right)
+{\cal C}_4\left(\zeta_2\tilde{\zeta}_3+\zeta_3\tilde{\zeta}_2\right)
\right]
\bigg\}\,,
\label{mesq}
\end{eqnarray}
where the newly--introduced unpolarized squared amplitude  
$\overline{\left|{\cal M}\right|^2_0}^{\,\prime}$ is given by
\begin{eqnarray}
\overline{\left|{\cal M}\right|^2_0}^{\,\prime}
= \overline{\left|{\cal M}\right|^2_0}+\beta^2\sin^4\Theta\,A_{\rm cont}^2 \,.
\end{eqnarray}
The second term above comes from the continuum contributions with opposite
photon helicities. The parameters  $\{\zeta_i\}$ and $\{\tilde{\zeta}_i\}$ 
($i=1,2,3$) are the Stokes parameters of two photon beams, respectively,
which should be initially prepared. In Sect.~\ref{sec:polarized photon beam}
we will give a brief description of generating energetic photon beams and
controlling their polarizations through the Compton laser backscattering off
the electron or positron beams.\\

The observables ${\cal B}_i$ ($i=1,2,3$) in Eq.~(\ref{mesq}) are due to the 
interference of the continuum and Higgs--boson parts in the helicity amplitudes
${\cal M}^{(\lambda\lambda:\sigma\sigma)}$ of Eq.~(\ref{me}) and they are 
explicitly given by
\begin{eqnarray}
&& {\cal B}_1=\bigg( {\cal A}_1-\beta {\cal A}_2+{\cal A}_3 \bigg)
             \,\overline{\left|{\cal M}\right|^2_0}
             \,\bigg/{\overline{\left|{\cal M}\right|^2_0}}^{\,\prime}\,,
             \nonumber \\
&& {\cal B}_2=2\bigg\{ \left|A_\phi\right|^2
             (\beta^2 S_t^2+P_t^2)\,{\cal R}(S_\gamma P^*_\gamma)
            +A_{\rm cont}\left[\beta^2 S_t\,{\cal R}(A_\phi P_\gamma) 
            +P_t\,{\cal R}(A_\phi S_\gamma)\right]\bigg\}
            \,\bigg/{\overline{\left|{\cal M}\right|^2_0}}^{\,\prime}\,,
            \nonumber \\
&& {\cal B}_3= \bigg\{(-1+\beta^2-\beta^2\sin^4\Theta) A_{\rm cont}^2
            +(\beta^2 S_t^2+P_t^2)(\left|S_\gamma\right|^2 
	                          -\left|P_\gamma\right|^2)
             \left|A_\phi\right|^2 \nonumber \\ 
   &&{ } \hskip 0.8cm +2A_{\rm cont}\left[ \vphantom{ {\cal M}^{(++:++)} }
             \beta^2 S_t{\cal R}(A_\phi S_\gamma)
             -P_t{\cal R}(A_\phi P_\gamma)\right]\bigg\}
             \,\bigg/{\overline{\left|{\cal M}\right|^2_0}}^{\,\prime}\,.
\end{eqnarray}
The observable ${\cal B}_3$ is CP--even, while the other two observables
${\cal B}_{1,2}$ are CP--odd. We note that the latter CP--odd observable
${\cal B}_2$ corresponds to the so--called T--odd  triple product of 
one photon momentum and two photon polarization vectors. \\

On the other hand, the five additional observables ${\cal C}_i$ ($i=0$ to
4) are due to the interference between the helicity amplitudes with  
equal photon helicities and those with opposite photon helicities and
they are explicitly given by
\begin{eqnarray}
&& {\cal C}_0=2\beta^2\sin^2\Theta A_{\rm cont}^2 \,
    \bigg/{\overline{\left|{\cal M}\right|^2_0}}^{\,\prime}\,,\nonumber \\
&& {\cal C}_1=2\beta^2A_{\rm cont}\bigg\{S_t\,{\cal R}(A_\phi P_\gamma) 
   \bigg\} \,\bigg/{\overline{\left|{\cal M}\right|^2_0}}^{\,\prime}\,,
   \nonumber \\
&& {\cal C}_2=2\beta^2A_{\rm cont}\bigg\{A_{\rm cont}
    +S_t\,{\cal R}(A_\phi S_\gamma) \bigg\}\,
    \bigg/{\overline{\left|{\cal M}\right|^2_0}}^{\,\prime}\,, \nonumber \\
&& {\cal C}_3=2\beta^2A_{\rm cont}\bigg\{S_t\,{\cal I}(A_\phi S_\gamma) 
    \bigg\}\,\bigg/{\overline{\left|{\cal M}\right|^2_0}}^{\,\prime}\,,
   \nonumber \\
&& {\cal C}_4=-2\beta^2A_{\rm cont}\bigg\{S_t\,{\cal I}(A_\phi P_\gamma) 
    \bigg\}\,\bigg/{\overline{\left|{\cal M}\right|^2_0}}^{\,\prime}\,.
\end{eqnarray}
Among these polarization asymmetries, the observables ${\cal C}_{0,2,3}$ are 
CP--even, while the observables ${\cal C}_{1,4}$ are CP--odd.

%%%%%%%%%%%%%%%%%%%%%%%%%%%%%%%%%%%%%%%%%%%%%%%%%%%%%%%%%%%%%%%%%%%%%%%%
\subsection{Top and anti--top quark polarizations}
%%%%%%%%%%%%%%%%%%%%%%%%%%%%%%%%%%%%%%%%%%%%%%%%%%%%%%%%%%%%%%%%%%%%%%%%

At asymptotically high energies the chirality conservation of the 
gauge--interactions leads to the dominance of $t \overline{t}$--pair
production with opposite helicity in the continuum amplitudes.
However, near the threshold, there is also substantial production 
of $t \overline{t}$--pairs of the same helicity.
The $t \overline{t}$ states with the same helicity transform to
each other under CP, so any asymmetry in their production rates
can provide a useful tool for studying CP violation. \\

Along with the initial two--photon polarizations we consider the final
polarization configuration with equal $t\bar{t}$ helicities to
construct the polarization asymmetry 
\begin{eqnarray}
\Delta=\frac{\overline{\left|{\cal M}\right|^2}\,(\sigma=\bar{\sigma}=+)-
             \overline{\left|{\cal M}\right|^2}\,(\sigma=\bar{\sigma}=-)}{
             \overline{\left|{\cal M}\right|^2_0}^\prime}\,.
\end{eqnarray}
Depending on the photon spin--spin correlations, the observable 
$\Delta$ is decomposed as follows:
\begin{eqnarray}
&& \Delta = {\cal D}_1\left(1+\zeta_2\tilde{\zeta}_2\right)
        +{\cal D}_2\left(\zeta_2+\tilde{\zeta}_2\right)
        +{\cal D}_3\left(\zeta_1\tilde{\zeta}_3+\zeta_3\tilde{\zeta}_1\right)
        -{\cal D}_4\left(\zeta_1\tilde{\zeta}_1-\zeta_3\tilde{\zeta}_3\right)
	 \nonumber \\
&&{ }\hskip 0.6cm 
        +\sin^2\Theta\bigg[{\cal E}_1\left(\zeta_1+\tilde{\zeta}_1\right)
        +{\cal E}_2\left(\zeta_3+\tilde{\zeta}_3\right)
        +{\cal E}_3\left(\zeta_1\tilde{\zeta}_2+\zeta_2\tilde{\zeta}_1\right)
        +{\cal E}_4\left(\zeta_2\tilde{\zeta}_3+\zeta_3\tilde{\zeta}_2\right)
        \bigg]\,.
\end{eqnarray}
The observables ${\cal D}_i$ ($i=1$ to $4$) are due to the interference of the
continuum and Higgs--boson parts in the helicity amplitudes
${\cal M}^{(\lambda\lambda:\sigma\sigma)}$ and they are explicitly 
given by
\begin{eqnarray}
&& {\cal D}_1=2\beta A_{\rm cont}\bigg\{-S_t\,{\cal I}(A_\phi P_\gamma)
                                       +P_t\,{\cal I}(A_\phi S_\gamma)\bigg\}
          \,\bigg/{\overline{\left|{\cal M}\right|^2_0}}^{\,\prime}\,,
          \nonumber \\
&& {\cal D}_2= 2\beta A_{\rm cont}\bigg\{A_{\rm cont} +
          \left[S_t\,{\cal R}(A_\phi S_\gamma)
	       +P_t\,{\cal R}(A_\phi P_\gamma)\right]\bigg\}
          \,\bigg/{\overline{\left|{\cal M}\right|^2_0}}^{\,\prime}\,,
	  \nonumber \\
&& {\cal D}_3= 2\beta A_{\rm cont}\bigg\{-S_t\,{\cal I}(A_\phi S_\gamma)
                                        +P_t\,{\cal I}(A_\phi P_\gamma)\bigg\}
          \,\bigg/{\overline{\left|{\cal M}\right|^2_0}}^{\,\prime}\,,
          \nonumber \\
&& {\cal D}_4= 2\beta A_{\rm cont}
          \bigg\{S_t\,{\cal I}(A_\phi P_\gamma)
	         +P_t\,{\cal I}(A_\phi S_\gamma)\bigg\}
          \,\bigg/{\overline{\left|{\cal M}\right|^2_0}}^{\,\prime}\,.
\end{eqnarray}
The observables ${\cal D}_{1,4}$ are CP--odd and the observables
${\cal D}_{2,3}$ are CP--even.\\

On the other hand, the four additional observables ${\cal E}_i$ ($i=1$ to
4) come from the interference between the helicity amplitudes with  
equal photon helicities and those with opposite photon helicities.
Their explicit forms are
\begin{eqnarray}
&& {\cal E}_1=2\beta A_{\rm cont}\bigg\{P_t\,{\cal I}(A_\phi P_\gamma)\bigg\}
             \,\bigg/{\overline{\left|{\cal M}\right|^2_0}}^{\,\prime}\,,
             \nonumber \\
&& {\cal E}_2=2\beta A_{\rm cont}\bigg\{P_t\,{\cal I}(A_\phi S_\gamma)\bigg\}
             \,\bigg/{\overline{\left|{\cal M}\right|^2_0}}^{\,\prime}\,,
             \nonumber \\
&& {\cal E}_3=-2\beta A_{\rm cont}\bigg\{P_t\,{\cal R}(A_\phi S_\gamma)\bigg\}
             \,\bigg/{\overline{\left|{\cal M}\right|^2_0}}^{\,\prime}\,,
             \nonumber \\
&& {\cal E}_4=2\beta A_{\rm cont}
             \bigg\{A_{\rm cont}+P_t\,{\cal R}(A_\phi P_\gamma)\bigg\}
             \,\bigg/{\overline{\left|{\cal M}\right|^2_0}}^{\,\prime}\,.
\end{eqnarray}
The observables ${\cal E}_{1,4}$ are CP--even and the observables
${\cal E}_{2,3}$ are CP--odd.

%%%%%%%%%%%%%%%%%%%%%%%%%%%%%%%%%%%%%%%%%%%%%%%%%%%%%%%%%%%%%%%%%%%%%%%%
\section{Complete measurements of the Higgs--boson CP property}
\label{sec:complete diagnosis}
%%%%%%%%%%%%%%%%%%%%%%%%%%%%%%%%%%%%%%%%%%%%%%%%%%%%%%%%%%%%%%%%%%%%%%%%

In order to completely determine the CP parity of the Higgs boson we
need to measure the following six quantities (see Eq.~(5)):
\begin{eqnarray}
\left\{m_\phi, \Gamma_\phi, S_\gamma, P_\gamma, S_t, P_t\right\}\,,
\end{eqnarray}
among which the one--loop induced $\gamma\gamma\phi$ form factors 
$S_\gamma$ and $P_\gamma$ are in general complex while the others
are real in the leading order. 
However, the helicity amplitudes are determined by
helicity--dependent multiplications of those quantities so that it
is necessary to measure the following quantities:
\begin{eqnarray}
&& S_t\,{\cal R}(A_\phi S_\gamma)\,,\  \
   S_t\,{\cal R}(A_\phi P_\gamma)\,,\  \
   S_t\,{\cal I}(A_\phi S_\gamma)\,,\  \
   S_t\,{\cal I}(A_\phi P_\gamma)\,,\nonumber\\
&& P_t\,{\cal R}(A_\phi S_\gamma)\,,\  \
   P_t\,{\cal R}(A_\phi P_\gamma)\,,\  \
   P_t\,{\cal I}(A_\phi S_\gamma)\,,\  \
   P_t\,{\cal I}(A_\phi P_\gamma)\,.
\label{eq:quantity}
\end{eqnarray}
The above 8 quantities are not completely independent and satisfy, for example, 
the following relations at all $s$ ;
%the c.m. energies $\sqrt{s}$;
% HAHA
\begin{eqnarray}
P_t\,{\cal R}(A_\phi S_\gamma) \cdot S_t\,{\cal R}(A_\phi P_\gamma) 
   &=&P_t\,{\cal R}(A_\phi P_\gamma) \cdot S_t\,{\cal R}(A_\phi S_\gamma)\,,
\nonumber \\[1mm]
P_t\,{\cal I}(A_\phi S_\gamma) \cdot S_t\,{\cal I}(A_\phi P_\gamma)
  &=& P_t\,{\cal I}(A_\phi P_\gamma) \cdot S_t\,{\cal I}(A_\phi S_\gamma)\,.
\label{eq:relation}
\end{eqnarray}
On the other hand, in principle 22 observables are available as shown 
in the previous section; ${\overline{\left|{\cal M}\right|^2_0}}$, 
${\overline{\left|{\cal M}\right|^2_0}}^{\,\prime}$,
4 ${\cal A}$'s, 3 ${\cal B}$'s, 5 ${\cal C}$'s, 
4 ${\cal D}$'s and 4 ${\cal E}$'s. 
Therefore, it is expected that the Higgs--boson parameters are
completely determined and that they are over--constrained. \\

Assuming that each observable is measured with a reasonable efficiency,
we provide a straightforward procedure to determine the quantities listed 
in Eq.~(\ref{eq:quantity}): 
\begin{itemize}
\item[{(1)}] The first four quantities in Eq.~(\ref{eq:quantity})
      can be determined directly through four observables 
      ${\cal C}_i$ ($i=1$ to 4) even without measuring top and anti-top
      helicities. 
%      Of course, for the sake of consistency, the determined 
%      quantities should satisfy several constraints from the observables  
%      ${\cal A}_0$, ${\cal A}_2$ and so on.

\item[{(2)}] The remaining four quantities in Eq.~(\ref{eq:quantity})
      can be determined directly through four observables 
      ${\cal E}_i$ ($i=1$ to $4$).
%      with the constraints from 
%      ${\cal A}_3$ and so forth. 
%%      In this case, the measurements of the top
%%      and anti-top polarizations are crucial.

\item[{(3)}] The constraints of Eq.~(\ref{eq:relation}), the observables
      ${\cal D}_i$ ($i=1$ to 4), and also ${\cal A}_i$ ($i=0$ to 3) and
      ${\cal B}_i$ ($i=1$ to 3) can be used to test and improve the 
      above measurements.

%      In addition, the observables ${\cal D}_i$ ($i=1$ to 4), which 
%      involve both the form factors $S_t$ and $P_t$ can be used as four 
%      consistency relations.
\end{itemize}
To recapitulate, the Higgs--exchange contribution can be completely determined 
by a judicious use of photon polarizations and $t\bar{t}$ helicity
measurements.

%%%%%%%%%%%%%%%%%%%%%%%%%%%%%%%%%%%%%%%%%%%%%%%%%%%%%%%%%%%%%%%%%%%%%%%%
\section{Polarized high energy photon beams}
\label{sec:polarized photon beam}
%%%%%%%%%%%%%%%%%%%%%%%%%%%%%%%%%%%%%%%%%%%%%%%%%%%%%%%%%%%%%%%%%%%%%%%%

The observed cross section is a convoluted one of the parton--level cross
section with a (polarized) $\gamma\gamma$ luminosity function
describing the spread of the $\gamma\gamma$ collision energy.
A detailed study of the possible luminosity and polarization distributions at 
future $\gamma\gamma$ colliders has been performed by the simulation 
program CAIN \cite{CAIN}. However, these quantities are strongly dependent on
the machine design of the colliders. Thus we adopt an ideal situation
of the beam conversion that the photon beam is generated by the tree--level
formula of the Compton backward--scattering and that the effect of the
finite scattering angle is negligible \cite{compton}.\\

High energy colliding beams of polarized photons can be generated by
Compton backscattering of polarized laser light on (polarized) 
electron/positron bunches of $e^+e^-$ linear colliders\footnote{In the present
work the electron and positron beams are assumed to be unpolarized.
It is however straightforward to take into account polarized electron 
and positron beams.}.
The polarization transfer from the laser light to the high energy photons is
described by three Stokes parameters $\zeta_{1,2,3}$; $\zeta_2$ is the
degree of circular polarization and $\{\zeta_3,\zeta_1\}$ the degree of
linear polarization transverse and normal to the plane defined by the
electron direction and the direction of the maximal linear polarization
of the initial laser light. Explicitly, the Stokes parameters take the 
form \cite{compton}:
\begin{eqnarray}
\zeta_1= \frac{f_3(y)}{f_0(y)}\,P_t\,\sin{2\kappa}\,,\qquad
\zeta_2=-\frac{f_2(y)}{f_0(y)}\,P_c\,,\qquad
\zeta_3= \frac{f_3(y)}{f_0(y)}\,P_t\,\cos{2\kappa}\,,
\label{stokes}
\end{eqnarray}
where $y$ is the energy fraction of the back--scattered photon with respect to
the initial electron energy $E_e$, $\{P_c, P_t\}$ are the degrees of 
circular and transverse polarization of the initial laser light, and
$\kappa$ is the azimuthal angle between the directions of initial photon 
and its maximum linear polarization.  Similar relations can be obtained 
for the Stokes parameters $\tilde{\zeta}$ of the opposite high energy photon
beam by replacing $(P_c, P_t, \kappa)$ with $(\tilde{P}_c, \tilde{P}_t, 
-\tilde{\kappa})$.  The functions $f_0$, $f_2$, and $f_3$ determining the
photon energy spectrum and the Stokes parameters are given by
\begin{eqnarray}
&& f_0(y)=\frac{1}{1-y}+1-y-4r(1-r) \,, \nonumber\\ 
&& f_2(y)=(2r-1)\left(\frac{1}{1-y}+1-y \right)\,, \nonumber\\
&& f_3(y)=2r^2 \,, 
\end{eqnarray}
with $r=y/x(1-y)$ and 
\begin{eqnarray}
x = \frac{4\,E_e\omega_0}{m_e^2}\approx 
15.4\,\Biggl( \frac{E_e}{ {\rm TeV}} \Biggr)
\Biggl( \frac{\omega_0}{ {\rm eV}} \Biggr)
\end{eqnarray}
for the initial laser energy $\omega_0$.
We note from Eq.~(\ref{stokes}) that the linear 
polarization of the high energy photon beam is proportional to $P_t$ whereas 
the circular polarization is proportional to $P_c$. Thus it is
necessary to have both circularly and linearly polarized initial laser
beams so as to measure all the polarization asymmetries ${\cal B}_i$'s and
${\cal C}_i$'s through the distribution (\ref{mesq}).\\

After folding the luminosity spectra of two photon beams, the event rate of 
the process $\gamma\gamma \rightarrow t\overline{t}$ is given by
{\small
\begin{eqnarray}
\frac{{\rm d}^3\,N}{{\rm d}\tau \,{\rm d}\Phi\,{\rm d}\cos\Theta}
&=&\frac{{\rm d}L_{\gamma\gamma}}{{\rm d}\tau }
   \frac{{\rm d}\hat{\sigma}_0}{{\rm d}\cos\Theta}
%   \frac{m^4_{\phi}}{2\pi^2}\,\hat{\sigma}_0
   \bigg\{1+\langle 22\rangle_\tau \, P_c\tilde{P}_c 
  -\langle 02\rangle_\tau \left(P_c+\tilde{P}_c\right)\,{\cal B}_1\nonumber\\
%&&{ }\hskip 0.5cm 
&&
  +\langle 33\rangle_\tau\, P_t\tilde{P}_t
   \bigg[\sin2(\kappa-\tilde{\kappa})\,{\cal B}_2
        +\cos2(\kappa-\tilde{\kappa})\,{\cal B}_3\bigg]\nonumber\\
%&&{ }\hskip 0.5cm 
&&
  -\sin^2\Theta\,\bigg[
   \left(\langle 22\rangle_\tau\, P_c \tilde{P}_c 
        -\langle 33\rangle_\tau\,
         P_t \tilde{P}_t \cos 2\kappa \cos 2\tilde{\kappa}\right)
	 \,{\cal C}_0\nonumber\\
%&&{ }\hskip 0.5cm
&&
  -\langle 03\rangle_\tau
   \left(\{P_t\sin 2\kappa-\tilde{P}_t\sin 2\tilde{\kappa}\}\,{\cal C}_1
        +\{P_t\cos 2\kappa+\tilde{P}_t\cos 2\tilde{\kappa}\}\,{\cal C}_2
   \right)\nonumber\\
%&&{ }\hskip 0.5cm
&&
  +\langle 23\rangle_\tau
   \left(\{P_t\tilde{P}_c\sin 2\kappa-\tilde{P}_tP_c\sin 2\tilde{\kappa}\}
         \,{\cal C}_3
        +\!\{P_t\tilde{P}_c\cos 2\kappa+\tilde{P}_tP_c\cos 2\tilde{\kappa}\}
	 \,{\cal C}_4
   \right)\bigg]\bigg\}\,,
\label{eq:fold}
\end{eqnarray}
}
where $\Phi$ is an azimuthal angle to be identifiable with $\kappa$, 
%the function $\frac{{\rm d}L_{\gamma\gamma}}{{\rm d}\tau}$  is
the function ${\rm d}L_{\gamma\gamma}/{\rm d}\tau$  is
the two--photon luminosity function depending
on the details such as the $e$-$\gamma$ 
conversion factor and the shape of the electron/positron bunches
\cite{compton}, and $\tau \equiv s/s_{ee}$.
The differential cross--section is then given by
\begin{eqnarray}
\frac{{\rm d}\hat{\sigma}_0}{{\rm d}\cos\Theta}=
\frac{\beta}{32\pi s}\, 
\overline{\left|{\cal M}\right|^2_0}^{\,\prime}\,.
\label{eq:dsigma}
\end{eqnarray}
The correlation ratios $\langle ij\rangle_\tau$ ($i,j=0$ to 3) 
are defined as
\begin{eqnarray}
\langle ij \rangle_\tau\equiv\frac{\langle f_i*f_j\rangle_\tau}{
                              \langle f_0*f_0\rangle_\tau}\,,
\end{eqnarray}
where the correlation function $\langle f_i*f_j\rangle_\tau$ is given by
the integrated function 
\begin{eqnarray}
\langle f_i*f_j\rangle_\tau=
 \int^{y_{\rm max}}_{\tau/y_{\rm max}}\frac{{\rm d}y}{y}\,f_i(y)f_j(\tau/y)
    \,,
\end{eqnarray}
with $y_{\rm max}=x/(1+x)$. The difference
$\kappa-\tilde{\kappa}$ of two azimuthal angles $\kappa$ and $\tilde{\kappa}$
is independent of the azimuthal angle $\Phi$ while each of them is linearly 
dependent on the angle $\Phi$. This implies that the measurements of the
observables ${\cal C}_i$ require the reconstruction of the scattering plane,
which can be done statistically.\\

\begin{figure}
 \begin{center}
%\hspace*{-5cm}
\epsfig{figure=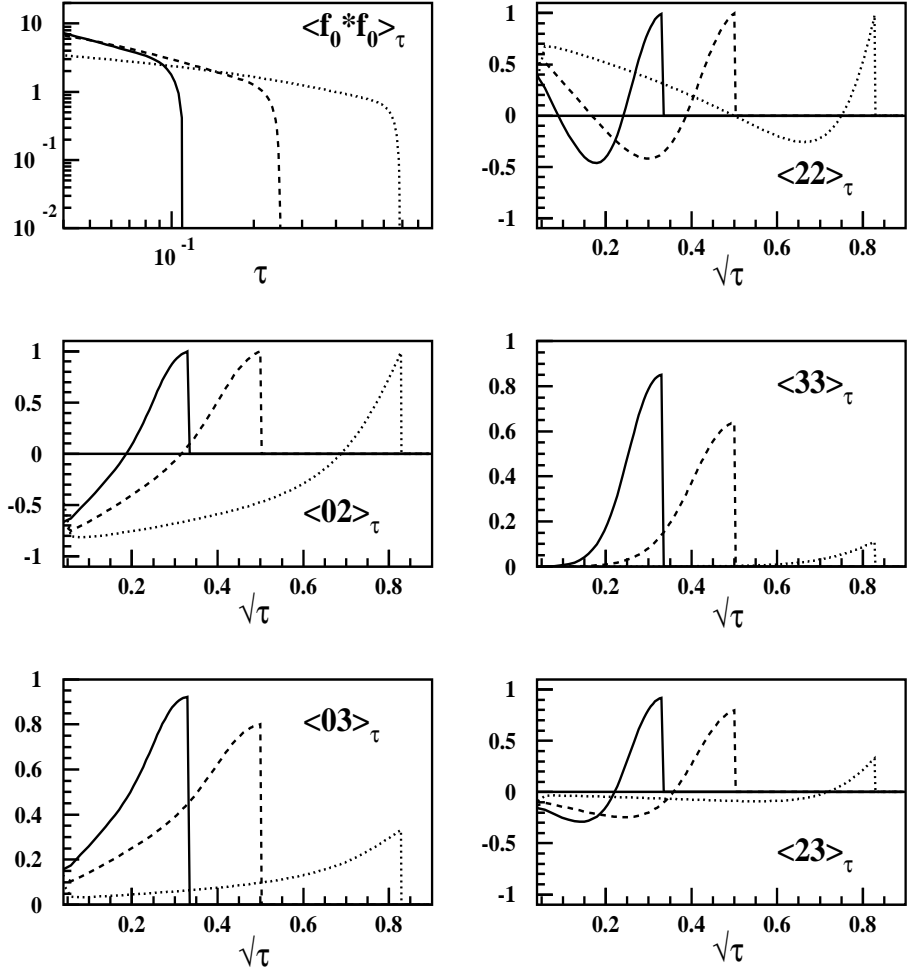,width=14cm,height=15cm}
%  \begin{picture}(140,180)
%   \put(-90,0){\epsfig{file=fig1.eps,width=17.5cm,height=8.3cm}}
%  \end{picture}
 \end{center}
\vskip -0.5cm
\caption{\it
          The unpolarized correlation function $\langle f_0*f_0\rangle_\tau$ 
          and the five ratios $\langle ij\rangle_\tau$ of the correlation 
	  functions for three $x$ values; $x=0.5$ (solid line), 
	  $1.0$ (dashed line), and $4.83$ (dotted line).}
\label{fig:fig1}
\end{figure}

Figure 1 shows the unpolarized correlation function  
$\langle f_0*f_0\rangle_\tau$ and the five correlation ratios 
$\{\langle 02\rangle_\tau,\langle 03\rangle_\tau,\langle 22\rangle_\tau,
\langle 23\rangle_\tau,\langle 33\rangle_\tau \}$ 
appearing in Eq.~(\ref{eq:fold}) for $x=0.5$ (solid line), 
$1.0$ (dashed line), and $4.83$ (dotted line). For a larger value of $x$, 
the correlation function $\langle f_0 * f_0 \rangle_\tau$, to which 
${\rm d}L_{\gamma\gamma}/{\rm d}\tau$ is proportional in the ideal 
situation of the beam conversion, becomes more flat and the 
maximally--obtainable photon energy fraction becomes 
closer to the electron beam energy. Exploiting this feature appropriately
could facilitate Higgs--boson searches at the photon collider.
The five figures for the correlation ratios clearly show
that the maximal sensitivity to each polarization asymmetry of 
${\cal B}_i$ and ${\cal C}_i$ can be acquired near the maximal value of  
$\tau=y_{\rm max}^2$. Therefore, once the Higgs--boson mass is 
known, one can obtain the maximal sensitivities by tuning the initial electron 
energy to be 
\begin{eqnarray}
E_e=\left(\frac{1+x}{2x}\right) m_{\phi} \,.
\end{eqnarray}
On the other hand, the ratios $\langle 33 \rangle_\tau$, $\langle 03 
\rangle_\tau$ and  $\langle 23 \rangle_\tau$ are larger for 
a smaller value of $x$ and for a given $x$ the maximum value of the ratio 
$\langle 33 \rangle_\tau$ is given by
\begin{eqnarray}
{\langle 33\rangle_\tau}_{\rm max}
=\left[\frac{2(1+x)}{1+(1+x)^2}\right]^2\,.
\end{eqnarray}
Consequently, it is necessary to take a small $x$ and a high $E_e$ 
by changing the laser beam energy $\omega_0$ so as to acquire the highest
sensitivity to CP violation in the neutral Higgs sector.

%%%%%%%%%%%%%%%%%%%%%%%%%%%%%%%%%%%%%%%%%%%%%%%%%%%%%%%%%%%%%%
\section{An example in the MSSM}
\label{sec:example}
%%%%%%%%%%%%%%%%%%%%%%%%%%%%%%%%%%%%%%%%%%%%%%%%%%%%%%%%%%%%%%

The MSSM Higgs sector constitutes a typical two--Higgs-doublet model 
in which CP violation can be induced at the loop level from the stop and 
sbottom sectors through the complex trilinear parameters $A_{t,b}$ and the
higgsino mass parameter $\mu$ \cite{EXCP,CDL}. Although there exist three 
neutral Higgs bosons, we consider only the heaviest Higgs boson so as 
to estimate the unpolarized parton--level cross section and the polarization 
asymmetries, which will allow us to completely determine the CP property 
of the Higgs boson following the procedure described in 
Sect.~\ref{sec:complete diagnosis}.

\begin{table}[\hbt]
\caption{\label{tab:table1}
{\it The mass and width $\{m_\phi, \Gamma_\phi\}$ and the four form factors 
     $\{S_\gamma, P_\gamma, S_t, P_t\}$ of the heaviest MSSM Higgs boson 
     for the parameter set (\ref{eq:para}) and $\tan\beta=3, 10$.}}
\begin{center}
\begin{tabular}{|c||c|c|c|c|c|c|}\hline
$\tan\beta$ & $m_\phi$ [GeV] & $\Gamma_\phi$ [GeV] & $S_\gamma$ & 
$P_\gamma$  & $S_t$ & $P_t$ \\ \hline\hline
3  & 500 & 1.9 & $-1.3-1.2\,i$   & $-0.51+1.1\,i$   & 0.33 &  0.15 \\ \hline
10 & 500 & 1.1 & $-0.39-0.35\,i$ & $-0.06+0.14\,i$ & 0.11 &  0.02 \\ 
\hline
\end{tabular}
\end{center}
\end{table}

In the MSSM the form factors $S_\gamma$ and $P_\gamma$, which describe
the coupling of the Higgs boson to two photons, have the loop contributions 
from the bottom and top quarks, the charged Higgs boson, the $W$ boson,
and the lighter top and bottom squarks as well as other charged particles 
such as charginos and heavier top and bottom squarks. The contributions 
from the charginos and heavier top and bottom squarks are neglected in 
the present work by taking them to be heavy.
% and those from the heavier sfermions can be
% neglected due to their very large masses. 
For our numerical example based on the work \cite{CDL}, we  
assume a universal trilinear parameter $A=A_t=A_b$ and 
take the physically--invariant phase $\Phi_{A\mu}=\Phi_A+\Phi_\mu$ to be 
$\pi/2$, leading to the (almost) maximal CP violation. 
Then, we take for the remaining dimensionful parameters the parameter 
set:
\begin{eqnarray}
|A|         = 1.0\,{\rm TeV}\,,\ \
|\mu|       = 2.0\,{\rm TeV}\,,\ \
M^2_{\tilde{Q}_{_L}}= M^2_{\tilde{t}_{_R}}= M^2_{\tilde{b}_{_R}}= 
(0.5\,{\rm TeV})^2\,, \ \
M_{H^{\pm}}=0.5\,{\rm TeV}
\label{eq:para}
\end{eqnarray}
where $M^2_{\tilde{Q}_{_L}}$, $M^2_{\tilde{t}_{_R}}$, and 
$M^2_{\tilde{b}_{_R}}$ are the soft SUSY breaking top/bottom squark masses 
squared  and $M_{H^{\pm}}$ is the charged Higgs boson mass.  In addition, we
take two values of $\tan\beta$, $\tan\beta=3$ and 10, 
so as to obtain a crude estimate of
the dependence of the form factors on $\tan\beta$. For the above MSSM 
parameters, the mass, the width, and the four form factors, 
$\{m_\phi,\Gamma_\phi, S_\gamma, P_\gamma, S_t, P_t\}$, 
of the heaviest MSSM Higgs boson on the mass pole
$\sqrt{s}=m_\phi$ are presented in Table \ref{tab:table1}.
Several comments on our results are in order:
\begin{itemize}
 \item The Higgs--boson width is reduced for large $\tan\beta$. This is
       due to the suppression of the dominant partial decay width 
       $\Gamma(\phi \rightarrow t\bar{t})$.
 \item The absolute values of all the form factors are very small 
       for large $\tan\beta$, leading to a strong suppression of the 
       Higgs--boson contribution to the process $\gamma\gamma\rightarrow
       t\bar{t}$. In particular, the `pseudoscalar' couplings, $P_t$ and
       $P_\gamma$, are very small, which implies an (almost) CP--even
       heavy Higgs boson.
\end{itemize}
One natural consequence from the $\tan\beta$ dependence of the form factors
is that all the CP--odd polarization asymmetries are strongly suppressed for 
large $\tan\beta$.\\

%It is a non--trivial job to experimentally measure the scattering angle 
%$\Theta$ of the final $t\bar{t}$ system through the decay products of the
%$t$ and $\bar{t}$ event by event. In this light, 
We integrate the polarized distributions over the angular variables  
so as to obtain the unpolarized parton--level cross section 
$\hat{\sigma}_0$ and the averaged polarization asymmetries:
%
% \begin{eqnarray}
% &&\overline{{\cal A}_i}\equiv
%    \frac{\int{\cal A}_i \overline{\left|{\cal M}\right|^2_0}\,
%          {\rm d}\cos\Theta}{\int\overline{\left|{\cal M}\right|^2_0}\,
%  	 {\rm d}\cos\Theta}, \hspace{0.5 cm}
%  \overline{{\cal B}_i}\equiv
%   \frac{\int{\cal B}_i \overline{\left|{\cal M}\right|^2_0}^{\,\prime}\,
%         {\rm d}\cos\Theta}{\int\overline{\left|{\cal M}
%	 \right|^2_0}^{\,\prime}\, {\rm d}\cos\Theta}\,, \hspace{0.5 cm}
%  \overline{{\cal D}_i}\equiv
%   \frac{\int{\cal D}_i \overline{\left|{\cal M}\right|^2_0}^{\,\prime}\,
%         {\rm d}\cos\Theta}{\int\overline{\left|{\cal M}
%	 \right|^2_0}^{\,\prime}\,{\rm d}\cos\Theta}\,, \nonumber \\ 
%&&
%\hspace{1.5 cm}
%\overline{{\cal C}_i}\equiv
%\frac{\int{\cal C}_i \overline{\left|{\cal M}\right|^2_0}^{\,\prime}
%\sin^2\Theta\,{\rm d}\cos\Theta} 
%{\int\overline{\left|{\cal M}\right|^2_0}^{\,\prime}\,{\rm d}\cos\Theta}\,,
%\hspace{0.5 cm}
%\overline{{\cal E}_i}\equiv
%\frac{\int{\cal E}_i \overline{\left|{\cal M}\right|^2_0}^{\,\prime}
%\sin^2\Theta\,{\rm d}\cos\Theta} 
%{\int\overline{\left|{\cal M}\right|^2_0}^{\,\prime}\,{\rm d}\cos\Theta}\,.
%\end{eqnarray}
%
\begin{eqnarray}
\overline{{\cal A}_i}\equiv \langle {\cal A}_i \rangle\,,\quad
  \overline{{\cal B}_i}\equiv \langle {\cal B}_i \rangle^\prime\,,\quad
\overline{{\cal C}_i}\equiv \langle {\cal C}_i\sin^2\Theta 
  \rangle^\prime\,,\quad 
  \overline{{\cal D}_i}\equiv \langle {\cal D}_i \rangle^\prime\,,
  \quad 
  \overline{{\cal E}_i}\equiv \langle {\cal E}_i\sin^2\Theta 
  \rangle^\prime\,,
\end{eqnarray} 
where $\langle X \rangle$ ($\langle X \rangle^\prime$) denotes the average 
over the distribution $\overline{\left|{\cal M}\right|^2_0}$ ( 
$\overline{\left|{\cal M}\right|^2_0}^{\,\prime}$).
We do not present the polarized distributions folded with the photon
luminosity spectrum explicitly because those distributions can be obtained
in a rather straightforward way from the parton--level cross sections. 
We do not take into account the QCD radiative corrections either, but 
for the details we refer to Ref.~\cite{SDGZ}.\\

\begin{table}[\hbt]
\caption{\label{tab:barobs1}
{\it The parton--level cross section and polarization 
     asymmetries $\overline{{\cal A}_{\mbox{ }}}$'s, which are constructed 
     with equal 
     photon and top--pair helicities, in the MSSM
     parameter set (\ref{eq:para}) for the CP phase $\Phi_{A\mu}=\pi/2$. 
     Each sign $\pm$ in the square brackets is for the CP--parity of the 
     observable.}}
\begin{center}
% \vspace{1.0cm}
\begin{tabular}{|c||c|c|c|c|c|}\hline
        &  &  &  &  &   \\[-4mm]
$\tan\beta$ & $\hat{\sigma}_0[+]$ &
              $\overline{{\cal A}_0}[+]$ & $\overline{{\cal A}_1}[-]$ &
	      $\overline{{\cal A}_2}[-]$ & 
	      $\overline{{\cal A}_3}[-]$ \\ \hline\hline
 3          &   0.88 pb           & $0.45$ & $ 0.13$ & $-0.17$  & $0.26$ \\ 
              \hline
 10         &   0.62 pb           & $0.91$ & $0.00$  & $-0.02$  & $0.03$ \\
	      \hline
\end{tabular}
\end{center}
\end{table}

Table~\ref{tab:barobs1} shows the parton--level unpolarized cross section 
$\hat{\sigma}_0$ and polarization asymmetries 
$\overline{{\cal A}_{\mbox{ }}}$'s, 
which are  constructed with equal photon helicities and equal $t\bar{t}$
helicities, in the MSSM parameter set (\ref{eq:para}) for the CP phase 
$\Phi_{A\mu}=\pi/2$. 
Each sign $\pm$ in the square brackets is for the CP--parity 
of the observable.  Note that the CP-odd observables $\overline{{\cal A}_1}$, 
$\overline{{\cal A}_2}$ and $\overline{{\cal A}_3}$ are significantly 
suppressed for $\tan\beta=10$ as expected.\\  

\begin{table}[\hbt]
\caption{\label{tab:barobs2}
{\it The values of the polarization asymmetries $\{\overline{{\cal B}_i}$ 
     and $\overline{{\cal C}_i}\}$, which are constructed with general 
     two--photon spin correlations, in the MSSM parameter set 
     (\ref{eq:para}) for the CP phase $\Phi=\pi/2$. 
     Each sign $\pm$ in the square brackets is for the CP--parity of the 
     observable.}}
\begin{center}
% \vspace{1.0cm}
\begin{tabular}{|c||c|c|c||c|c|c|c|c|}\hline
        &  &  &  &  & &  &  &  \\[-4mm]
$\tan\beta$ & $\overline{{\cal B}_1}[-]$ & $\overline{{\cal B}_2}[-]$ &
	      $\overline{{\cal B}_3}[+]$ &
              $\overline{{\cal C}_0}[+]$ & $\overline{{\cal C}_1}[-]$ &
	      $\overline{{\cal C}_2}[+]$ & 
	      $\overline{{\cal C}_3}[+]$ &
	      $\overline{{\cal C}_4}[-]$\\\hline\hline
 3          &  $ 0.46$ & $-0.27$ & $-0.60$ &
	       $ 0.17$ & $ 0.13$ & $ 0.09$ & $0.17$ & $ 0.06$\\ \hline
 10         &  $ 0.03$ & $ 0.00$ & $-0.47$ &
	       $ 0.24$ & $ 0.01$ & $ 0.30$ & $0.04$ & $ 0.00$ \\
	       \hline
\end{tabular}
\end{center}
\end{table}
\begin{table}[\hbt]
\caption{\label{tab:barobs3}
{\it The polarization asymmetries $\overline{{\cal D}}$'s and 
     $\overline{{\cal E}}$'s, which are constructed with general two--photon 
     spin correlations and equal top--pair helicities, in the MSSM parameter 
     set (\ref{eq:para}).  The re-phasing--invariant phase $\Phi_{A\mu}$ is 
     taken to be $\pi/2$ and only the interference between the continuum 
     and the heaviest Higgs boson in the MSSM is taken into account.
     The signature $\pm$ in the square brackets 
     denotes the CP--parity of the corresponding observable.}}
\begin{center}
% \vspace{1.0cm}
\begin{tabular}{|c||c|c|c|c||c|c|c|c|}\hline
  & & & & & & & & \\[-4mm]
 $\tan\beta$  & $\overline{{\cal D}_1}[-]$   & $\overline{{\cal D}_2}[+]$   &
 $\overline{{\cal D}_3}[+]$   &
                $\overline{{\cal D}_4}[-]$ & 
                $\overline{{\cal E}_1}[+]$   & $\overline{{\cal E}_2}[-]$   &
		$\overline{{\cal E}_3}[-]$   &
		$\overline{{\cal E}_4}[+]$\\ \hline\hline
 3            & $ 0.32$ & $ 0.41$ & $-0.45$ & $ 0.03$ &
                $-0.04$ & $ 0.10$ & $ 0.09$ & $ 0.40$ \\\hline
 10           & $ 0.03$ & $ 0.80$ & $-0.09$ & $ 0.00$ &
                $-0.00$ & $ 0.01$ & $ 0.01$ & $0.46$\\ \hline
\end{tabular}
\end{center}
\end{table}

In Tables~\ref{tab:barobs2} and \ref{tab:barobs3}, we show
the polarization asymmetries $\{\overline{{\cal B}_i}$,  
$\overline{{\cal C}_i}\}$ constructed with general two--photon 
spin correlations, and  the polarization asymmetries $\overline{{\cal D}}$'s 
and $\overline{{\cal E}}$'s with general two--photon 
spin correlations and equal $t$ and $\bar{t}$ helicities, 
in the MSSM parameter set (\ref{eq:para}). 
The re-phasing--invariant phase $\Phi_{A\mu}$ is taken to 
be $\pi/2$ and each signature $\pm$ in the square brackets is for the 
CP--parity of the observable as in Table~\ref{tab:barobs1}. The CP--odd
polarization asymmetries are sizable for $\tan\beta=3$, but they are
strongly suppressed for $\tan\beta=10$. 
In addition, the CP--even 
observables which are dependent only on the products $S_\gamma 
P_\gamma$ and $S_tP_t$ of the form factors are also suppressed for large 
$\tan\beta$. 
In particular, we note that the polarization asymmetry ${\cal E}_1$ is 
determined by only the CP-odd (`pseudoscalar')
form factors $P_\gamma$ and
$P_t$ so that its strong suppression implies an almost CP--even Higgs boson.\\

%%%%%%%%%%%%%%%%%%%%%%
\section{Conclusions}
\label{sec:conclusion}
%%%%%%%%%%%%%%%%%%%%%%

In this article, we have studied the effects of a neutral Higgs boson 
without definite CP--parity in the process 
$\gamma\gamma\rightarrow t\bar{t}$ in a model--independent way.\\

We have found that the interference between the Higgs--boson--exchange 
and the continuum amplitudes with polarized photon beams 
and with the top and anti--top helicity measurements enables us 
to determine the CP property of a neutral Higgs boson completely
even in CP non--invariant theories. 
We have classified the physical observables such as the unpolarized
cross sections and polarization asymmetries depending on the polarization
configuration of the initial two--photon beams and the helicity
configuration of the final $t\bar{t}$ system. 
As an demonstration of the procedure,
the contribution of the heaviest neutral Higgs boson 
in the MSSM has been investigated quantitatively. \\
%In addition, we have described the generation mechanism of circularly
%and linearly polarized photon beams through the Compton laser backscattering
%off unpolarized electron/positron beams.\\

Certainly, the precision with which the Higgs--boson contribution is
determined depends on the background processes as well as the efficiencies
for controlling beam polarization and measuring the $t$ and $\bar{t}$ 
helicities. Nevertheless, the algorithm presented in the present
work with more than 20 polarization observables will be helpful in 
determining the $s$--channel Higgs contribution to the process 
$\gamma\gamma\rightarrow t\bar{t}$ efficiently even in the CP non--invariant
theories.

%%%%%%%%%%%%%%%%%%%%%%%%%%%%%
\section*{Acknowledgements}
%%%%%%%%%%%%%%%%%%%%%%%%%%%%%

S.Y.C. wishes to acknowledge financial support of the 1997 Sughak program of
the Korea Research Foundation.

\vskip 0.3cm

%

%

%\newpage
\setcounter{equation}{0}
\renewcommand{\theequation}{A\arabic{equation}}

%\end{multicols}

\end{document}